\documentclass[12pt,preprint]{aastex}
\usepackage{emulateapj5}

\begin{document}

\noindent
{\it Dissertation Summary}

\begin{center}

%%% put title of your dissertation in following line:

\title{\large \bf Quantitative Spectroscopy of Supergiants }

\end{center}

%%% Your name and current address below:

\author{ Norbert Przybilla }

\affil{Institute for Astronomy, 2680 Woodlawn Drive, Honolulu, HI 96822, USA}

\begingroup

\parindent=1cm

%%% supply the following information:

\begin{center}

Electronic mail: norbert@ifa.hawaii.edu

Thesis work conducted at: Ludwig-Maximilians-Universit\"at M\"unchen

Ph.D. Thesis directed by: R.P.~Kudritzki/R. Bender;~Ph.D. Degree awarded: 
June 2002

%%% do not change following line:

{\it Received \underline{\hskip 5cm}}

\end{center}

\endgroup

%%% fill in appropriate keywords from the list on the PASP web site
%% (http://pasp.phys.uvic.ca):

\keywords{line: formation -- stars: abundances -- stars: evolution -- 
stars: fundamental parameters -- supergiants -- galaxies: abundances}

%%%place the text of your Dissertation Summary here:

Blue supergiants of spectral types B and A are the visually brightest stars
in spiral and irregular galaxies, with their most luminous members (at
$M_V$\,$\simeq$\,$-$10) outshining entire dwarf galaxies. This characteristic 
allows us to use them as probes to study the Local Universe in great detail. 
In principle, already the existing large telescopes and 
instrumentation facilitate quantitative spectroscopy of these objects as far as 
the Virgo and Fornax clusters of galaxies. Beyond their challenging
stellar atmospheres and opportunities for testing sophisticated non-LTE physics they offer 
numerous applications to modern astrophysics.
Quantitative spectroscopy of supergiants will contribute to improve our
understanding of massive star evolution. Galactic abundance gradients and
abundance patterns, as will be obtained from studies of large ensembles
of supergiants in our own and other galaxies, will foster the understanding of 
galactochemical evolution. Finally, they are promising independent indicators
for calibrating the extragalactic distance scale, by application of the wind 
momentum--luminosity and the flux-weighted gravity--luminosity relationships 
(Kudritzki et al.~1999; Kudritzki, Bresolin \& Przybilla~2003).

In view of this large potential, the objective of this thesis is to improve the 
status of quantitative spectroscopy of BA-type supergiants and to provide first 
applications on a sample of Galactic and extragalactic targets. 
It is shown that among the model atmospheres available {\em at present}, 
the best suited for analyses of supergiants are 
line-blanketed classical LTE models. 
An investigation of the impact of various parameters such as helium abundance
and line blanketing on the atmospheric structure shows
that for the most luminous objects an accurate treatment of these parameters 
is essential for a quantitative analysis, whereas the less luminous 
supergiants react less sensitive. Spectrum synthesis is used to model the line spectra. 
It is the only technique capable of providing analyses of spectra 
of different qualities from low to high resolution, and able to cope
with heavy line blending at a broad range of signal-to-noise ratios.
The stellar parameters are determined from purely spectroscopic indicators,
from temperature and gravity sensitive ionization equilibria and the Balmer
line wings. Elemental abundances are derived by modelling
individual spectral features.
Several ten-thousand spectral lines from 28 chemical species are included 
in the line formation, allowing almost the entire observed spectra to be 
reproduced. Non-LTE effects become important in 
blue supergiants, where a strong radiation field at low particle densities
favours deviations from LTE. Comprehensive model atoms are therefore 
constructed for C\,{\sc i/ii}, N\,{\sc i/ii}, O\,{\sc i} and Mg\,{\sc i/ii} 
in order to determine non-LTE level populations 
(Przybilla, Butler \& Kudritzki~2001; Przybilla \& Butler~2001; 
Przybilla et al.~2000,~2001)
Highly accurate radiative and
collisional atomic data recently determined for astrophysical and fusion 
research using the $R$-matrix method in the close-coupling approximation are 
incorporated. In addition, model atoms for H,
He\,{\sc i}, O\,{\sc ii}, S\,{\sc ii/iii}, Ti\,{\sc ii} and Fe\,{\sc ii} are
adopted from the literature, the atomic data being updated to more modern
values in some cases. Thus, an improved treatment for the main elements of 
astrophysical interest is achieved.

Extensive testing of the atomic data is performed for
the nearby bright main sequence standard Vega (A0\,V), at well-determined 
stellar parameters and atmospheric structure. A
high-resolution and low-noise 
spectrum with large wavelength coverage from the visual to the near-IR
is used for this purpose. Further tests are performed for the Galactic 
supergiants
$\eta$\,Leo, HD\,111613, HD\,92207 and 
$\beta$\,Ori, with similar high-quality spectra
in order to study the non-LTE effects across the parameter space. 
Accurate and consistent stellar parameters are derived for these.
Non-LTE ionization equilibria of several elements~-- typically 
N\,{\sc i/ii}, O\,{\sc i/ii}, Mg\,{\sc i/ii} and S\,{\sc ii/iii} -- agree
simultaneously, provided a realistic treatment of line-blocking is used.
These parameters also constitute important input data for further studies of
the stellar winds of these objects.
Accounting for non-LTE reduces the random errors and removes systematic
trends in the analyses. In particular, the improved treatment of electron
collisions largely removes long-standing discrepancies in analyses of lines
from different spin systems of a given ion.
The computed non-LTE line profiles fit the 
observations well for the different species at a given elemental 
abundance. In the parameter range covered, all lines from He\,{\sc i},
C\,{\sc i/ii}, N\,{\sc i/ii}, O\,{\sc i/ii} and S\,{\sc ii/iii} are
significantly strengthened by non-LTE effects; Mg\,{\sc ii} remains almost
unaffected, except for the strongest lines; non-LTE weakening is found for
the lines of Mg\,{\sc i}, Ti\,{\sc ii} and Fe\,{\sc ii} in supergiants.
The nature of the non-LTE effects is
investigated: non-LTE strengthening generally occurs in conjunction with a
strong overpopulation of metastable energy levels, while non-LTE weakening
is due to overionization of minor ionic species. In extreme cases, as for
several strong lines of N\,{\sc i} and O\,{\sc i}, non-LTE abundance
corrections up to a factor of 50 to 100 are found, whereas typical mean
non-LTE abundance corrections for the diagnostic lines are within a factor
of up to~3. Estimates of the systematic uncertainties in the non-LTE abundance
analysis of CNO and Mg are provided. Accounting for these and random
errors, it is shown that absolute elemental abundances can
be derived with accuracies of $\sim$0.1 to 0.25\,dex (1$\sigma$-uncertainties), 
depending on the element, in contrast to $\sim$0.2--0.3\,dex (only random 
contribution) typically achieved in previous
studies. The statistical significance of the analyses 
is also largely improved due to the large wavelength coverage of the
spectra, with many lines per element being available. 
In addition, hitherto unaccounted effects on metal
line strengths are found from the consistent treatment of microturbulence 
in the non-LTE computations and line formation.

The abundance analysis for Vega confirmes its status as a mild $\lambda$
Bootis star: the light elements, CNO, show an underabundance of
$\sim$0.25\,dex when compared to the solar composition, while the heavier
elements are depleted by $\sim$0.55\,dex.
All four Galactic supergiants have metallicities close to solar. The
non-LTE abundances for individual heavier elements in a star cluster around 
a mean offset to the solar composition, whereas in a classical LTE analysis
misleading `abundance patterns' are seen. In particular, LTE
analyses tend to overestimate the abundances of the $\alpha$--process elements
and to underestimate the Iron Group abundances; this systematic effect
strengthens with increasing luminosity. It might be suspected that other 
elements with an atomic structure comparable to the species
investigated are susceptible to similar non-LTE mechanisms;
these need to be quantified in future studies. The increasing scatter
of individual abundances around a mean value with increasing stellar effective
temperature and luminosity can be interpreted in terms of neglected effects
in the atmospheric modelling. Non-LTE effects on the atmospheric
structure become more pronounced at higher temperature, while sphericity 
and hydrodynamical outflow velocity fields are noticeable only for very
luminous objects.

A similar analysis is performed on high-resolution spectra of supergiants
in nearby Local Group galaxies, although at lower signal-to-noise ratios.
%(Przybilla et al.~2002b).
VLT/UVES and Keck/HIRES spectra are available for this 
(Venn et al.~2000,~2001).
One early A-type supergiant near the centre of the dwarf irregular galaxy
NGC\,6822 is found to be metal-poor by $\sim$0.55\,dex, confirming similar
values from the literature, both from stellar and H\,{\sc ii} region studies.
The non-LTE overionization of Ti\,{\sc ii}
shows a distinctive strengthening at this low metallicity.
Two objects of early-A spectral type are analysed in 
M\,31, both located at a galactocentric distance of
$\sim$12\,kpc. The objects have abundances compatible with and slightly above 
solar, also in good agreement with measurements from close-by H\,{\sc ii} regions
in M\,31.
The non-LTE effects follow the same trends as in the Galactic counterparts.

While indications are found that the abundances of the heavier
elements in the supergiants cluster around a mean value, distinctive
abundance patterns for the light elements are derived, with
enriched He and N, depleted C, and O being compatible with the heavier
element abundances. The predictions for chemical mixing from recent stellar 
evolution models,
accounting for mass-loss and rotation, can be verified: good agreement is
found for the He enrichment and the N/C ratios in the Galactic supergiants,
whereas the observed N/O ratios are lower than those from the models.
%(Meynet \& Maeder~2000)
With the presently achieved accuracy in the abundance determination,
different stages of stellar evolution -- first crossing of the
Hertzsprung-Russell diagram from the main sequence to the red vs. a
blue-loop scenario with first dredge-up abundances -- can be distinguished
with confidence. From the Galactic sample, $\eta$\,Leo is apparently 
undergoing a
blue-loop, while the other stars have directly evolved from the main sequence.
%(Przybilla~2002).%; Przybilla et al.~2003).
In the NGC\,6822 and M\,31 supergiants similar He enrichments and unchanged
O abundances are derived. However, as the diagnostic lines of the 
main indicators for mixing, C and N, are not covered by the available spectra,
further conclusions on their evolutionary status cannot be drawn at present.

Low-resolution spectra of supergiants in the Sculptor Group spiral galaxy
NGC\,300 and in NGC\,3621 in the field, at distances of~2.0 and 6.6\,Mpc --
well beyond the Local Group --,
were recently obtained with FORS1 on the VLT (Bresolin et al.~2001,~2002). In order to 
perform quantitative analyses of these stars, the applicability of the spectrum
synthesis technique is tested on the most luminous Galactic sample star,
HD\,92207, its spectrum being artificially degraded to FORS1 resolution.
It is shown that the stellar parameters can be determined with sufficient
accuracy from the spectral classification and the Balmer line strengths to
constrain the metallicity to $\pm$0.2\,dex. Also, individual abundances
for a few key elements of astrophysics can be determined at this resolution.
Two early A-type~supergiants in NGC\,3621 %(Przybilla et al.~2002a) 
and two similar objects in
NGC\,300 are studied, each among the most luminous stars in these galaxies.
The objects have metallicities of slightly sub-solar to
$\sim$0.2\,$\times$\,solar, typically, in concordance with expectation from 
their position in their host galaxies. 
The stellar metallicities agree with literature data on 
abundance gradients for these galaxies, as derived from H\,{\sc ii} regions;
the reddening of the objects is compatible with being due to the Galactic
foreground. Analyses of stars at such distances are performed for the first 
time. 

%%% if there are references, place below, following standard reference style,
%% or to save space you may use "in-line" references within the text and delete
%% the lines below:

%\begin{references}

%\reference{  XXXX, Y.Z. 1998, ApJ, 234, 456}  

%\end{references}

\end{document}